# Towards Cross-Surface Immersion Using Low Cost Multi-Sensory Output Cues to Support Proxemics and Kinesics Across Heterogeneous Systems

**Rajiv Khadka**
University of Wyoming, 3DIA Lab
1000 E. University Ave,
Dept. 3315, Laramie, WY 82071
rkhadka@uwyo.edu

**James Money**
Idaho National Laboratory
Applied Visualization Laboratory
995 University Blvd.,
Idaho Falls, ID 83401
James.Money@inl.gov

**Amy Banic**
University of Wyoming, 3DIA Lab
Idaho National Laboratory, AVL
1000 E. University Ave,
Dept. 3315, Laramie, WY 82071
abanic@cs.uwyo.edu

## Abstract
Collaboration in immersive systems can be achieved by using an immersive display system (i.e. CAVE and Head-Mounted Display), but how do we communicate immersion cross-surface for low immersive displays, such as desktops, tablets, and smartphones? In this paper, we present a discussion of proxemics and kinesics to support based on observation of physical collaboration. We present our research agenda to investigate low-cost multi-sensory output cues to communicate proxemics and kinesics aspects cross-surface. Doing so may increase the level of presence, co-presence, and immersion, and improve the effectiveness of collaboration cross-surface.

## ACM Classification Keywords
H.5.m. Information interfaces and presentation (e.g., HCI): Miscellaneous.

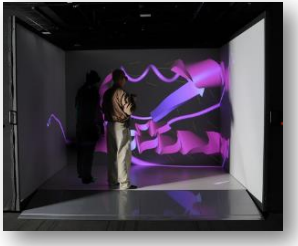

Figure 1: Example of an Immersive System (CAVE Automated Virtual Environment).

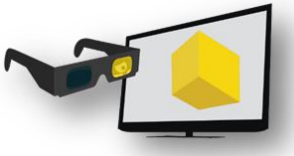

Figure 2: Example of a low immersive system.

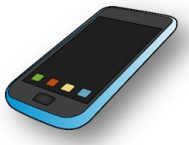

Figure 3: Example of a non-immersive system.

## Introduction

Collaborative Virtual Environments (CVEs) provides a shared virtual workspace where people from different geographical locations can meet and interact with each other. They can share and work on 3D objects/models to achieve common goals [3]. CVEs are increasingly being used to support collaborative work between co-located or geographically remote collaborators. Collaborators can work between different levels of immersive display systems, from highly immersive (CAVE and Head-Mounted Display) to low immersive (desktop, tablet, and smartphone) display systems [1]. However, there are difficulties in providing a realistic user experience and immersive quality to users in low immersive systems when collaborating with others who are interacting in an immersive environment in a shared networked virtual environment. Immersive display systems have a high degree of presence in comparison with a display system that does not support aspects to create immersion [1]. Increased sense of presence and co-presence in a system helps to increase the effectiveness of the collaborative work [4], transfer of skills to real world [12], and improves skills in collaborative manipulation [15]. Only a few have conducted design and evaluation of collaboration using cross-display collaborative virtual environments [19]. CVEs for collaboration provide a strong platform for learning, understanding and evaluating complex spatial information. While working in CVEs, a decrease in immersion due to the field of view, the field of regard, display size, tracking available, etc. leads to a decrease in co-presence. Since co-presence has been shown to enhance collaboration performance [5], we would like to explore alternative methods to increase co-presence among systems with lower levels of immersion. In this paper, we present aspects in proxemics and kinesics which should be supported cross-surface due to importance for effective collaboration. We present our research agenda for designing and investigating low-cost multi-sensory output cues to be used for a low immersive display system in order to communicate proxemics and kinesics from immersive systems to non-immersive systems. Our goal is to design and investigate low-cost proxemics cue techniques as a way to increase co-presence such that these cues help to improve collaborative workflow across heterogeneous systems for a variety of applications.

## Proxemics

Proxemics refers to the study of how space and distance influence communication [11]. There are four larger zones of proxemics: public, social, personal, and intimate distance. The amount of space defined for each of these zones varies across cultures and social organizations. Public and social zones refer to space that is four or more feet away from our body, and the communication that typically occurs in these zones is formal and not intimate [11]. Communication that occurs in the social zone, which is four to twelve feet away from our body, is typically in the context of a professional or casual interaction, but not intimate or public. Personal and intimate zones refer to the space that starts at our physical body and extends a short distance (for US Americans it is about four feet). These zones are reserved for friends, close acquaintants, and significant others. The intimate zone, reserved for only the closest friends, family, and romantic/intimate partners, extends within about one to two feet of the body. With individuals entering this space, it is difficult for individuals to ignore others in this space. One additional smaller zone of proxemics is pericutaneous space [7]. This refers to a layer of space very near to and a feeling just prior to touching. Visual-tactile perceptive fields overlap in processing this space. For example, an individual might see a feather as not touching their skin but still experience the sensation of being tickled when it hovers just above their hand. Other examples include the blowing of wind, gusts of air, and the passage of heat. We are very interested in studying this space in the context of cross-surface interaction. Using this phenomenon, we plan to add feedback mechanisms to serve as cues to indicate nearness, distance, direction of human contact.

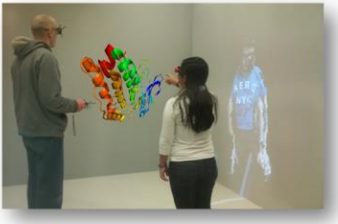

Figure 4: Collaboration in an Immersive Display System

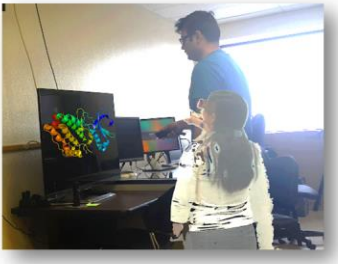

Figure 5: Collaboration in a non-immersive display system. Example shown is video avatar projected in the space.

### Kinesics
There are three main types of gestures: adaptors, emblems, and illustrators [11]. Adaptors are touching behaviors and movements that indicate internal states typically related to arousal or anxiety. Emblems are gestures that have a specific agreed-on meaning. Illustrators are the most common type of gesture and are used to illustrate the verbal message they accompany.

### Advances in Collaborative Virtual Environments
A highly immersive room-sized, fully dynamic real-time 3d scene capture and continuous-viewpoint head-tracked display on a life-sized tiled display wall was developed for symmetric collaboration (i.e. using same immersive display system across all remote participants) [17]. The system used abundant sensor information, cameras, high-quality displays for increased level of immersion. Another research work supporting symmetric collaboration incorporated three-way communication over a distributed shared workspace which was designed to support three channels of communication: person, reference, and task-space [20]. Researchers developed a highly immersive telepresence system for symmetric collaboration that allowed distributed group of collaborators to meet and share their 3D environment among each other easily [2]. More recent work involving collaboration across heterogeneous devices, such as supporting interaction across 3D workstation, desktops, and Personal Digital Assistants (PDA's), was conducted [20]. A data-centric design was used for synchronous collaboration. A recent approach for collaboration used large interactive virtual spaces i.e. high-resolution wall-sized displays and CAVE for remote collaboration across heterogeneous collaborative virtual environment [9]. This research identified that sense of presence in a virtual environment helps in understanding users' interaction capabilities which eventually helps to increase the workflow of the collaborative task. Other important aspects for collaboration are co-presence, social presence, spatial proximity, relationships, and context, which have all been shown to range from higher to lower for face-to-face interaction, immersive CVEs, video-CVEs, and standard CVEs respectively [14,18]. The face-to-face condition was experienced as being significantly helpful in collaborative task [13]. Proxemics have effects on the collaborator's behavior, communication, and social interactions. Proxemics also enhance the collaborator's ability to work together and collaborate easily. [8,10]. So, with higher proximity and more cues of awareness, we can increase copresence and which then should increase efficiency in collaboration tasks.

### Proxemics and Kinetics Found in Physical Collaboration with Single-Display
We conducted a study to investigate how users collaborate for data exploration and analysis in the same physical environment or co-located space (see fig. 1), using an immersive display system at Idaho National Laboratory. We collected action and observation data on participants exploring data environments and performing analysis tasks while collaborating with other participants in the co-located space. This study was designed to help us understand how the participants would interact and use the immersive virtual environment while collaborating their work. Our analysis of the data collected describes the collaborative behavior and interactions exhibited by the participants in a co-located immersive space while interacting with an immersive application.

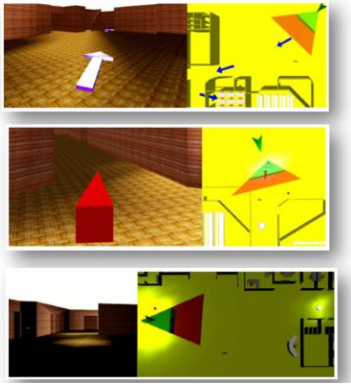

Figure 6: Visual Cues for proxemics. Photo from [6]

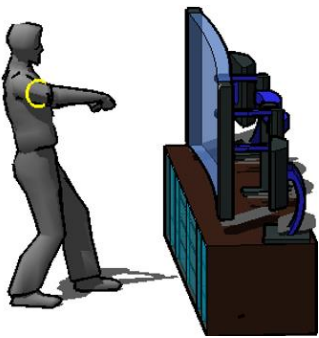

Figure 7: Participant wearing vibrotactile belt in their arms using low immersive display system during collaboration.

*Proxemics to Support*
While working with remote collaborative virtual environments, sense of presence between collaborators can be preserved better through 3-dimensional video avatars [4]. However, current implementations do not necessarily preserve all the relationships between the user and environment as they would appear to be in the physical space of the immersive display. These types of relationships should be preserved when providing these representations. In the next section we outline our plan to help increase the user experience of the collaborators in relation to spatial direction, distance, and awareness of type of collaborator. Furthermore, there should be a distinction between touching, nearness to touch, within reach but not touching, and not in reach. The reason is that effective collaboration is correlated with the combination of the factors nearness and relationship of collaborators. Use of audio as interaction techniques may also help to increase the level of user experience collaborators experiences. Spatial audio should be preserved cross-surface to facilitate collaborative communication.

Spatial audio helps with the challenge of multiple collaborators talking in a CVE.

*Kinesics to Support*
Gestures that supported collaboration were broken down into most frequent and specific gestures used. A majority of the participants were trying to grab and touch data being displayed in the application. The next most frequently used gestures were those which included pointing at a specific area of the immersive application in which they want to collaborate and communicate with the co-located collaborators. This result shows that spatial relationships between users and each other, users and data, and users and the display are essential for communication and task completion for collaboration. So, when we are using heterogeneous collaborative virtual environment which includes low immersive display systems communicating the proximity play an essential part for effective collaboration. Spatial relationships of gestures between users, between users and environment, and between users and the display, need to be preserved and rescaled to map appropriately to the physical space of each individual collaborator. Representations need to be provided for the users' gestures, however, can be enhanced through visual interaction history trails to better illustrate the physical body to environment relationship. We preserve these relationships by providing cues in the environment.

## Research Agenda for Designing and Investigating Proxemics and Kinesics

In this section, we present our position on designing and investigating low-cost proxemic cues techniques for low immersive display systems. We have identified that there is a need to add feedback mechanism to serve as cues to indicate nearness, distance, and direction from human contact. There is also a need to communicate gesture action, direction, and content. In this section, we discuss and present our research agenda to design and investigate proxemic and kinesic cues to enhance collaboration cross-surface.

*Levels of Proxemics*
In our prior work we found that known collaborators, the distance is closer for more effective collaboration. For collaborators who do not know each other well, further distance makes for more effective collaboration. We seek to answer the question, what if we could break these conventions in cross-surface? For example, for known collaborators always communicate nearness, even when far away. For unknown collaborators, communicate in public space, even if in the virtual environment their locations are rather close. There are multiple levels of proxemics, as defined in the section on 'Proxemics'. We shall investigate the use of visual, vibrotactile feedback, and spatial audio to determine which and what variations more appropriately communicate nearness, touch potential, distance, direction, and collaborator relationship from an immersive environment to a non-immersive environment. In the following sections, we detail more specifically what we will investigate for each.

*Visual Cues*
Based on data from our user study, using natural gestures is the most effective method of interaction in an immersive virtual environment. Spatial relationships of gestures between users, between users and data, and between users and the display, need to be preserved and rescaled to map appropriately to the physical space of each individual collaborator. So, we want to use visual cues i.e. glyphs (to be used as a control condition) and a light source (see fig. 5) with projectors to provide and increase the sense of presence in the appropriate location while using a low immersive display system. A set of visual solutions we will investigate include cues on the edges of the display frame, small projector-based solutions in the environment, and glassware with small LEDs in the periphery. Aspects to communicate include varying the type of representation for collaborator relationship, the size and/or color based on proximity, and location based on direction. To communicate pericutaneous space, we will need to investigate subtle ubiquitous changes in visual information. These visual solutions may help individual collaborators to communicate their directional and distance information in respect to another collaborator more effectively during collaboration.

Glyphs: Each collaborator will be assigned an individual glyph. Each glyph will have a distinct color and a name of collaborator on top of the glyph. So, that each glyph can be distinguished easily. With the increase in the nearness of the collaborator, the size of the glyph will also be increased and otherwise. With the increase in the nearness of the collaborator, the distance of the glyph will also decrease. The glyph will be pointed to show the point of view of the collaborator.

Small Pico Projectors: In low resolution, each collaborator will be assigned a separate color of light source. For high resolution, the avatar of the collaborator will appear. With the increase in the nearness of collaborator, the intensity of the light will also increase and decrease as the distance of the collaborator increases. The light source will be projected depending upon the direction of the position of the other collaborator.

Glassware with LEDs: The intensity of blinking will increase or decrease with the increase or decrease in the distance of the collaborators. The left led or right led will blink in respect to a drastic change in the direction of the collaborator. A constant glow will emit based on the direction of the collaborator and slightly increase or decrease in intensity based on distance.

*Vibrotactile Cues*
There has been an increasing amount of work using vibrotactile belts in the virtual environment to communicate a sense of presence and touch. We also plan to investigate the use vibrotactile belts, shoulder and arm-wear, or within the device itself. Collaborators to wear vibrotactile belts on both arms (see fig. 6). One solution is that when a collaborator is moving near to the collaborator, the vibration of the belt will increase or decrease in frequency or strength accordingly to inform the collaborator the direction and distance in respect to the another collaborator. Low vibration or locational vibration may distinguish between pericutaneous space and less intimate space. These vibrotactile cues may help to increase the sense of presence and copresence while a collaborator is using a low immersive display system and will be compared to our visual solutions.

*Temperature and Pressure Cues*
We will also explore the production and release of temperature and pressure as cues. We will vary the increase and decrease as well as the frequency applied.

*Spatial Audio Cues*
Spatial audio should be preserved across the immersive visualization environments to facilitate collaborative communication. However, in this section we discuss in addition to spatial audio itself, to use as a means to communicate other proxemics and kinesics information. It will be a challenge to be able to balance any use of audio with actual audio from collaborators. These may be in the form of non-verbal sounds or more ubiquitous rising and falling of consistent music in the background. We will compare the differences among visual, vibrotactile and spatial audio cues.

## Summary and Conclusion

In this paper, our research agenda to design and investigate low-cost multi-sensory output cues as a way to increase co-presence such that these cues help to improve collaborative workflow cross-surface from immersive systems to non-immersive systems. Spatial relationships between users and environment, users, and users with the display technology are important and need to be preserved across heterogeneous environments among collaborators whether those are high immersive or low immersive display system. And the preservation of interaction, gestures, communication mechanisms, and spatial relationships should be adapted cross-surface for simulating immersion.